\documentclass{aastex}
\usepackage{emulateapj5}
\submitted{Submitted to ApJL, July 1st 2002}

\usepackage{epsfig}
\slugcomment{}
\shorttitle{}
\shortauthors{}

\begin{document}

\title{The magnification of SN 1997\lowercase{ff}, 
the farthest known Supernova}

\author{Narciso Ben\'\i tez\altaffilmark{1}, 
        Adam Riess\altaffilmark{2}, 
	Peter Nugent\altaffilmark{3}, 
	Mark Dickinson\altaffilmark{2},
        Ryan Chornock\altaffilmark{4}, and
        Alexei V. Filippenko\altaffilmark{4}
}

\altaffiltext{1}{Department of Physics and Astronomy, The
Johns Hopkins University, 3400 N. Charles Street, Baltimore MD 21218, 
txitxo@pha.jhu.edu}
\altaffiltext{2}{Space Telescope Science Institute, 3700 San Martin Drive, 
Baltimore, MD 21218}
\altaffiltext{3}{Lawrence Berkeley National Laboratory, Berkeley, CA 94720}
\altaffiltext{4}{Department of Astronomy, 601 Campbell Hall, University of 
California, Berkeley, CA  94720-3411}

\begin{abstract}

With a redshift of $z \approx 1.7$, SN 1997ff is the most distant type Ia
supernova discovered so far. This SN is close to several bright, $z = 0.6-0.9$
galaxies, and we consider the effects of lensing by those objects on the
magnitude of SN 1997ff. We estimate their velocity dispersions using the
Tully-Fisher and Faber-Jackson relations corrected for evolution effects, and
calculate, applying the multiple-plane lensing formalism, that SN 1997ff is
magnified by $0.34 \pm 0.12$ mag. Due to the spatial configuration of the
foreground galaxies, the shear from individual lenses partially cancels out,
and the total distortion induced on the host galaxy is considerably smaller
than that produced by a single lens having the same magnification.  After
correction for lensing, the revised distance to SN 1997ff is $m-M = 45.5$ mag,
which improves the agreement with the $\Omega_M =0.35,\Omega_\Lambda = 0.65$
cosmology expected from lower-redshift SNe~Ia, and is inconsistent at the $\sim
3\sigma$ confidence level with a uniform gray dust model or a simple evolution
model. 

\end{abstract}


\section{Introduction}

One of the strongest lines of evidence for an accelerating universe is the
observation that \( z \approx 0.5 \) type Ia supernovae (SNe~Ia) are \( \sim
0.25 \) mag fainter than predicted by an \( \Omega_\Lambda =0 \) cosmology
(Riess et al. 1998; Perlmutter et al. 1999; Riess 2000). However, there are
several effects such as gray dust (Aguirre 1999) or luminosity evolution
(Drell, Loredo, \& Wasserman 2000) that could mimic the effect of a
cosmological constant. 

  Riess et al. (2001, hereafter R2001) have recently concluded that SN 1997ff,
one of the two SNe discovered by Gilliland, Nugent, \& Phillips (1999) in the
Hubble Deep Field North (HDFN; Williams et al. 1996), has a redshift of \( z
\approx 1.7, \) making it the highest redshift probable SN~Ia identified so
far. (The SN~Ia classification is based on the nature of the host galaxy, an
evolved, red elliptical; also the observed colors and temporal evolution of SN
1997ff are most consistent with those of SNe~Ia.)  This object is particularly
interesting because at \( z > 1 \) the preceding epoch of deceleration causes
objects to appear brighter than a persistence of astrophysical effects invoked
as alternative hypotheses to explain the SN~Ia faintness at \( z \approx 0.5
\). R2001 measured a distance modulus for SN 1997ff which is more than a
magnitude brighter ($1.25 \pm 0.32$ mag brighter assuming the spectroscopic
indication of $z = 1.755$) than the predictions from uniform gray dust or toy
evolution models, disfavoring these alternatives to a cosmological constant.

Lewis \& Ibata (2001) stated that SN 1997ff could be magnified by a factor of
$ \sim 1.4$ by two ``elliptical systems" close to its line of sight.  R2001
anticipated a magnification of $0.3$ mag from these same sources, in good
agreement with Lewis \& Ibata.  As we see below, these are in fact two spiral
galaxies (\#709 and \#720 in Table 1) which magnify SN 1997ff by only $\sim
0.2$ mag, about half of the value estimated by Lewis \& Ibata, and close to the
upper limit inferred by R2001 for these two galaxies from the induced shear on
the host-galaxy shape and orientation.

M{\" o}rtsell, Gunnarsson, \& Goobar (2001) also tackled this problem using
rest-frame $B$-band and ultraviolet luminosities to infer the masses of the
lenses. After performing numerous ray-tracing simulations, they concluded that
the range of possible magnifications was too large to obtain a firm value for
the distance modulus of SN 1997ff based on our current knowledge of the lensing
galaxies.

Here we use the superb photometric and positional information provided by the
HDFN to estimate velocity dispersions for the galaxies lensing SN 1997ff using
the Tully-Fisher (TF) and Faber-Jackson relations, and correcting for evolution
effects.  With this information, we calculate, using the multiple-plane lensing
formalism (Schneider, Ehlers, \& Falco 1992, hereafter SEF), the magnification
of SN 1997ff and the shear induced on the host galaxy, which is fully
compatible with its expected intrinsic shape.
 
  The structure of the paper is as follows. Section 2 describes the environment
of SN 1997ff, and identifies the main lensing foreground galaxies. The
multiple-plane lens formalism is discussed in \S~3, while \S~4 presents our
estimates of the magnification and shear. Section 5 summarizes our main results
and conclusions.

\section{The lenses}

Figure 1 shows the positions of the galaxies around SN 1997ff with \(
I_{814W} < 27 \) mag and $r < 15\arcsec$. The coordinates, ID number, and \(
I_{814W} \) magnitudes are taken from the Fern\'andez-Soto, Lanzetta, \& Yahil
(1999) catalog and photometric redshifts have been obtained using the BPZ package 
(Ben\'\i tez 2000, http://acs.pha.jhu.edu/\~txitxo/bpz.html). Table 1 lists the 
characteristics of the galaxies contributing significantly to the magnification 
of SN 1997ff.
The $B$ and $I$ absolute magnitudes $M_B$ and $M_I$ have been estimated from
the \( I_{814W} \) and {\it HST}/NICMOS F110W and F160W magnitudes,
respectively. This was done by generating K-corrections based on their spectral
type, drawn from the 4 main Coleman, Wu, \& Weedman (1980) galaxy types and two
Kinney et al. (1996) starbursts; the K-corrections are  
$\lesssim 0.4$ mag in the case of the four strongest lenses.

%Note that although spectral type does not necessarily 
%correspond to morphological type, visual inspection shows that in this case both 
%classifications broadly agree. 

The circular velocities $V_c$ for the late-type galaxies in Table 1 have been
estimated using the recent results of Ziegler et al. (2002), which measure the
$B$-band TF slope and normalization for $R < 23$ mag, $0.5 < z < 1$ FORS deep field
galaxies --- quite similar, in statistical terms, to our lens sample. We assume a
relative error of $\delta V_c/V_c \approx 1/3$ estimated from the scatter
around the best fit in Figure 3 of Ziegler et al. (2002). These errors are
propagated in the calculation of the Einstein radius, etc., and account for most
of the uncertainty in the final value of the magnification. As a cross-check, we 
have also estimated the expected values of $V_c$ usingthe $I$-band TF
relation, as determined by Giovanelli et al. (1997): the results are very similar 
to those of the Ziegler et al. (2002) $B$-band TF.

%This seems to indicate that the rest-frame $I$ band is
%not as affected by recent star formation activity as bluer bands, and thus can
%also produce a reliable estimate of the total mass of the galaxies, at least up
%to $z \approx 0.6$. We nevertheless decide to use the results of Ziegler et al.
%since they match better the characteristics of our sample, but the coincidence
%between the two estimates, based on very different data sets and filters, adds
%robustness to the results quoted in Table 1.

  To determine their circular velocities with another method, we observed
galaxies \#709 and \#720 on 2001 May 29 and 30 UT with the Echellette
Spectrograph and Imager (ESI; Sheinis et al. 2002) on the Keck-II 10-m
telescope.  A slit width of $1''$ was chosen, which gave an instrumental
resolution of $\sim$ 70 km s$^{-1}$ (based on fits to the night-sky emission
lines). The spectrum of galaxy \#709 shows strong, single-component emission
lines with FWHM = 78 km s$^{-1}$ (corrected for the instrumental resolution).
The spectrum of galaxy \#720 exhibits double-peaked emission-line profiles,
with a peak-to-peak splitting of 200 km s$^{-1}$. After correcting for the
assumed inclination angles ($34^\circ$ for galaxy \#709 and $69^\circ$ for
galaxy \#720), we derive circular velocities of 70 and 110 km s$^{-1}$,
respectively. In contrast, the respective circular velocities expected
according to the TF relation are substantially higher, $\sim 136 \pm 48$ and
$\sim 175 \pm 60 $ km s$^{-1}$ (Table 1). Although both values are marginally
consistent at the $\sim 1.25\sigma$ level, these results, especially in 
the case of \#709, seem to support the conclusion of Gillespie \& van Zee 
(2002) that emission lines from the central regions of star-forming galaxies 
can underestimate the total circular velocity by a factor of $\sim 2$. 
Hence, we consider the circular velocities derived from the Keck data
to be lower limits on the true quantities, and we use the TF estimates and
their uncertainties in the calculations below. Taking the spectroscopic results
at their face value would yield $\Delta\mu = 0.07$ mag from the \#709--\#720 pair
alone, which can be considered to be the lower limit on the magnification of SN
1997ff. Note also that this would make the total magnification smaller (by
$\sim 0.16$ mag), and would further strengthen our cosmological conclusions
on gray dust or evolution models.

%The fact that the slits were not placed exactly along the major axis of the 
%galaxies (since the main purpose of the observations was to obtain the 
%redshift of the SN 1997ff host) also complicates the analysis and 
%interpretation of these results. 

For the two early-type galaxies, \#694 and \#653, we use the Faber-Jackson
relation between the halo velocity dispersion and the $B_J$-band luminosity as
defined by Kochanek (1994, 1996), $\sigma_{DM} = (225 \pm 22.5) (L/L_*)^4$, where
$L_*$ corresponds to $B_J = -19.9 + 5\log h$ mag (Efstathiou, Ellis, \& Peterson
1988) and $h = H_0/(100$ km s$^{-1}$ Mpc$^{-1}$). Owing to the redshift of these
galaxies, $z \approx 0.9$, we expect evolution effects to be considerable
(e.g., Treu et al. 2002; although see also Crampton et al. 2002). Keeton, Kochanek,
\& Falco (1998) studied the properties of a sample of 17 lensing galaxies with
$0.1 < z < 1$, and measured an evolution rate for the mass-to-light ratio of
$d\log(M/L_B)/dz \approx -0.6 \pm 0.1$ for the cosmological model considered
here.  After accounting for this effect, the velocity dispersions of galaxies
\#694 and \#653 become about half of their value in the absence of evolution.

\section{The lensing equations}

Because of the configuration of the lenses, placed at different redshifts, it is
necessary to use the multiple lens-plane formalism (SEF) to estimate the
magnification and shear on the SN 1997ff position. The magnification matrix for
$N$ lenses at, in general, different positions and redshifts, can be
represented as

\begin{equation}
A=I-\sum _{i=1}^{N}A_{i}, ~~~~~~~~~~~A_{j}=I-\sum _{i=1}^{j-1}\beta _{ij}U_{i}A_{i},
\end{equation}

\noindent
where \( A_{1}=I \) (the identity matrix), the coefficients \( \beta _{ij} \)
are distance ratios (\( \beta _{ij}=0 \) if all the galaxies are
on the sample plane), and \( U_{i} \) can be represented as 

\[
U_{i}=\kappa _{i}I+\gamma _{i}\left( \begin{array}{cc}
\cos 2\phi _{i} & \sin 2\phi _{i}\\
\sin 2\phi _{i} & -\cos 2\phi _{i}
\end{array}\right). \]

\noindent
In this equation \( \kappa _{i} \) and \( \gamma _{i} \) correspond to the
convergence and shear induced by each $i$-galaxy on the source in the absence
of other lenses. 
Assuming that the mass distribution of the galaxies can be
well represented by a singular isothermal sphere, we have

\[
\kappa _{i}=\frac{\theta _{ei}}{2\theta _{i}},~~~~~~\gamma _{i}=-\frac{\theta
 _{ei}}{2\theta _{i}}e^{i2\phi }. \]

\noindent
The Einstein radius \( \theta _{ei} \) can be represented as 
\( \theta _{ei}=4\pi (V_{c}/c)^{2}D_{is}/D_{s} \)
(Bartelmann \& Schneider 2001), where \( V_{c} \) is the circular velocity,
\( D_{is} \) the distance between the $i$-galaxy and the source,
and \( D_{s} \) is the distance between the source and the observer.
Finally, if the matrix \( A \) is symmetric, it can be represented
in the usual way as 

\[
A=\left( \begin{array}{cc}
1-\kappa -\gamma _{1} & -\gamma _{2}\\
-\gamma _{2} & \kappa 
\end{array}\right), \]

\noindent
where the total shear is \( \gamma =\gamma _{1}+i\gamma _{2} \) and
the total magnification is \( \mu =[(1-\kappa )^{2}+\gamma ^{2}]^{-1} \).

To describe the effects of lensing on the shape of a galaxy, we can
represent it with the complex ellipticity $\chi =
[(1-q^{2})/(1+q^{2})]e^{i2\phi }$, 
where \( q=a/b \) is the axis ratio, \( 0<q<1 \). Given a reduced
shear \( g=\gamma /(1-\kappa ) \), the intrinsic ellipticity of the
source can be recovered as (Schneider \& Seitz 1995)

\[
\chi _{s}=\frac{2g+\chi +g^{2}\chi ^{*}}{1+|g|^{2}+2Re(g\chi ^{*})}.
\]

\noindent
The above equations have been implemented as a Python module which can 
be obtained by contacting N. Ben\'\i tez.

%We will assume that the mass distribution of the galaxies can be
%well represented by a singular isothermal sphere and use the corresponding equations 
%from, for example, SEF.

%, we have
%
%\[
%\kappa _{i}=\frac{\theta _{ei}}{2\theta _{i}},~~~~~~\gamma _{i}=-\frac{\theta
% _{ei}}{2\theta _{i}}e^{i2\phi }. \]
%
%\noindent
%The Einstein radius \( \theta _{ei} \) can be represented as 
%\( \theta _{ei}=4\pi (V_{c}/c)^{2}D_{is}/D_{s} \)
%(Bartelmann \& Schneider 2001), where \( V_{c} \) is the circular velocity,
%\( D_{is} \) the distance between the $i$-galaxy and the source,
%and \( D_{s} \) is the distance between the source and the observer.
%Finally, if the matrix \( A \) is symmetric, it can be represented
%in the usual way as 

%\[
%A=\left( \begin{array}{cc}
%1-\kappa -\gamma _{1} & -\gamma _{2}\\
%-\gamma _{2} & \kappa 
%\end{array}\right), \]

%\noindent
%where the total shear is \( \gamma =\gamma _{1}+i\gamma _{2} \) and
%the total magnification is \( \mu =[(1-\kappa )^{2}+\gamma ^{2}]^{-1} \).

%To describe the effects of lensing on the shape of a galaxy, we can
%represent it with the complex ellipticity $\chi =
%[(1-q^{2})/(1+q^{2})]e^{i2\phi }$, 
%where \( q=a/b \) is the axis ratio, \( 0<q<1 \). Given a reduced
%shear \( g=\kappa /(1-\gamma ) \), the intrinsic ellipticity of the
%source can be recovered as (Schneider \& Seitz 1995)

%\[
%\chi _{s}=\frac{2g+\chi +g^{2}\chi ^{*}}{1+|g|^{2}+2Re(g\chi ^{*})}.
%\]

\noindent

\section{Results}

\subsection{The Magnification}

Using the above expressions, we find that the lensing correction to the
brightness of SN 1997ff is a magnification of $\Delta m_\mu = 0.34 \pm 
0.12$ mag, for an $\Omega_M=0.35$, $\Omega_\Lambda=0.65$ cosmology and 
using the filled-beam approximation to the angular distance. 
Table 1 shows the Einstein radius and magnification $\mu_i$ induced on the
position of SN 1997ff by each individual galaxy.
If we naively multiply all these magnifications, we arrive at a
value of $2.5 \log(\Pi_i \mu_i) = 0.31$, which represents only a $10\%$ 
difference from the multiple lens-plane formalism result.

%The values for a $\Omega_\Lambda=0.0$
%model are essentially the same. The reason for this is that the Ziegler et
%al. (2002) TF calibration, which defines the mass of the main lenses, is
%basically an empirical relation between the apparent magnitudes of $z \approx
%0.6$ galaxies and their velocity dispersions, and as such, is independent of
%the cosmological model.

\subsection{The Host Ellipticity}

R2001 estimated that, from the ellipticity and orientation of the host galaxy,
the probability of its being magnified by more than 0.34 mag by the galaxy
pair \#709 and \#720 (the two strongest lenses) is only 18\%. 
Our new result for the total shear is 
$\gamma = 0.084$, half of the value expected for an equivalent single isothermal
sphere, and along the direction perpendicular to the \#709 + \#720 pair.
Given this reduced shear, 
the host ellipticity is significantly less constraining in providing an
upper limit to the magnification; the 90\% confidence limits for the 
magnification are (0.086, 1.37). The cause of this reduced leverage of the
distortion constraint is that the shear vectors from the galaxies whose
contribution was not considered in R2001 partially cancel out.  The result is
that, unlike the magnification ($\sim 0.13$ mag), the distortion they
induce on the SN 1997ff host is very small, and slightly counteracts that
caused by the \#709 + \#720 pair, further reducing the total shear.

%This constraint
%was based on the fact that weak lensing by an isothermal sphere induces a shear
%of roughly $|\gamma| \approx (\mu-1)/2$, which would make the intrinsic shape
%of the host galaxy unusually elongated. 
%After applying the multiple lens-plane
%formalism, and taking into account the contributions from all the galaxies
%close to SN 1997ff, not just \#709 and \#720, we find that the total shear is

\subsection{Cosmological Constraints}

 R2001 measured a distance modulus of $m - M=45.15 \pm 0.32$ mag, assuming
$z_s = 1.755$. After correction for lensing, the new value is $m - M = 45.49 \pm
0.34$ mag, in excellent agreement with the expected value of $45.68$ mag for
the $\Omega_M=0.35, \Omega_\Lambda=0.65$ ($h=0.65$) universe expected from the
$z \approx 0.5$ population of SNe~Ia (Riess et al. 1998).

   Using the preceding calculation to correct the SN 1997ff distance modulus
for lensing, we can compare the result to toy models that invoke astrophysical
sources of dimming instead of a cosmological constant to match the observations
of SNe~Ia at $z \approx 0.5$.  Here we will assume the spectroscopic redshift
measurement of $z = 1.755$ for SN 1997ff to simplify the evaluation of the
astrophysical dimming expected at a specific redshift. The astrophysical
dimming described in R2001 would take the functional form of $0.3z$ mag
relative to an empty Universe.  At $z = 1.755$ this results in a predicted
distance modulus of $\sim$46.4 mag ($h = 0.65$).  The lensing-corrected distance
modulus of SN 1997ff is 45.49 mag, smaller (i.e., the SN is brighter) than the
astrophysical dimming model by 0.91 mag (reduced from 1.25 mag without the
lensing correction).  As discussed by R2001, the fit to the SN data yields
non-Gaussian and asymmetric uncertainties beyond the $2\sigma$ interval.
Returning to the fit from R2001 to the SN data (corrected by the slight
increase in the distance modulus error due to the lensing uncertainty) yields
confidence intervals in the direction of larger distances of $+1\sigma$,
$+2\sigma$, and $+3\sigma$ equal to 0.34, 0.67, and 0.90 mag, respectively.  We
therefore conclude that the simplest alternatives to an accelerating universe
appear to be inconsistent with SN 1997ff at the $3\sigma$ confidence level.
However, we caution that the astrophysical model of dimming is naive and before
tests of astrophysical sources can be considered robust, independent
confirmation of this result will be required from additional SNe~Ia at $z > 1$.

  Mannheim (2001) has shown that the observations of $z < 1$ SNe can be
accommodated within a conformal gravity cosmological model (Mannheim 1992) with
$q_0 = -0.37$, which predicts a distance modulus of $m-M = 46.11$ at $z =
1.755$. This agrees within $1.8\sigma$ with our new result, implying that more
data are necessary to discriminate between this hypothesis and a cosmological
constant.

\section{Discussion}

The magnification of SN 1997ff, \( \mu \approx 1.4 \), is high but compatible 
with the expectation from simulations. For instance, 
Barber et al. (2000) find \(\mu >1.4 \) for \( 5\% \) of the lines 
of sight to \( z=2 \). 
However, we would not attribute such magnification to a selection effect; the 
dispersion of observed magnitudes due to lensing is only $\sim$0.2 mag, no larger 
than the intrinsic dispersion of SNe~Ia, and it has been shown to cause negligible 
selection bias (Riess et al. 1998).
 
%A lensing simulation using the HDF-N galaxies shows that the line of sight to SN1997ff 
%is also in the $5\%$ magnification quartile in this field. 
%To check
%whether SN 1997ff occupies a privileged position, a Monte Carlo simulation is
%performed using the HDFN galaxies. Random positions are chosen within $10''$ of
%the field borders and the magnification at \( z=1.7 \) is calculated using all
%the foreground galaxies up to that redshift. The line of sight to SN 1997ff is
%indeed uncommonly magnified in the statistical sense; only \( 5\% \) of the
%HDFN positions have larger values of \( \mu . \) 
%As expected, due to the approach employed here the average magnification, \(
%<\mu >=1.06 \), is larger than 1, the minimum possible magnification in this
%scenario. This is comparable with our uncertainty in the magnification
%estimation, and given the small size of the HDFN, it is far from clear whether
%this would correspond to the average magnification over the entire sky. For
%example, there are indications that the HDFN has a factor \( \sim 2 \)
%overdensity of galaxies around \( z\approx 1 \) compared with the HDFS (Fontana
%et al. 2000).

  It is unlikely that lensing by large-scale structure could significantly
affect our results. Dalal et al. (2002) show that most of the scatter in the
magnification comes from scales of less than $1'$, corresponding to the range
considered here. The expected residual root-mean-square on larger scales is
$\sim 0.07$ mag, smaller than our uncertainty. It is noteworthy that by
combining an ``individual lenses'' approach like the one followed here, with a
shear analysis which estimates the large-scale magnification as proposed by
Dalal et al., it could be possible to significantly reduce the uncertainty in
supernova distances due to lensing. Dark haloes (without optically visible
galaxies) may be a possible problem with this approach, but so far there is
very little, if any, evidence for their existence. The major source of error in
our calculations is the dispersion in the TF and Faber-Jackson relations at
high redshift, but in the future improved calibrations using near-infrared data
should reduce this scatter considerably.

The amount of magnification we expect for SN 1997ff is similar to the values
determined by Lewis \& Ibata (2001) and R2001, and roughly agrees with the
results of M{\" o}rtsell et al. (2001) for realistic values of
their input parameters. However, our use of the multiple lens-plane formalism,
individual galaxy K-corrections, and TF and Faber-Jackson relationships,
corrected for evolution, provides an estimate that is more accurate and robust
than previous work.

In summary, SN 1997ff, with \( z\approx 1.7 \), is the highest redshift SN~Ia
discovered so far. It is shown here that gravitational lensing by nearby
foreground galaxies is likely to have magnified SN 1997ff by \( 0.34 \pm 0.12\)
mag. Due to the spatial configuration of the foreground galaxies, the shear
from the individual lenses partially cancels out, and the total distortion
induced on the host galaxy is less than half of that produced by a single lens
with the same magnification.  After correcting for lensing, and assuming $z_s =
1.755$, the distance modulus to SN 1997ff is $m-M = 45.49 \pm 0.34$ mag, in
better agreement with the $\Omega_M =0.35,\Omega_\Lambda =0.65$ cosmology
expected from lower-redshift SNe~Ia, but inconsistent with uniform gray dust or
simple evolution models as an explanation for the dimming of $z < 1$ SNe~Ia at
the $3\sigma$ confidence level. It is noteworthy that our new result also
agrees within $1.8\sigma$ with the predictions of conformal gravity
cosmological models (Mannheim 2001).

Since the study of high-z SNe~Ia cannot avoid
the effects of gravitational lensing, future use of SNe to constrain
cosmological parameters will need to consider and ultimately contend with the
effects of lensing.

\acknowledgements{}

The authors thank John Blakeslee for useful comments and suggestions.
N.B. acknowledges financial support from the NASA ACS grant. P.E.N.
acknowledges support from a NASA LTSA grant and by the Director, Office of
Science under U.S. Department of Energy Contract
No. DE-AC03-76SF00098. A.V.F. is grateful for NSF grant AST-9987438 and a
Guggenheim Foundation Fellowship. The Keck Observatory was made possible by the
generous financial support of the W. M. Keck Foundation.

\begin{deluxetable}{rlrrrrrrrcrrrr}
\tabletypesize{\scriptsize}
\tablecaption{Galaxies magnifying SN 1997ff ($\Omega_M=0.35$, $\Omega_\Lambda=0.65, h=0.65$)}
\tablewidth{0pt}
\tablehead{
\colhead{ID\tablenotemark{a}} & 
\colhead{HDF-ID\tablenotemark{b}} & 
\colhead{X\tablenotemark{c}} & 
\colhead{Y\tablenotemark{c}} & 
\colhead{R\tablenotemark{c}} &
\colhead{$z_{s}$\tablenotemark{d}} &
\colhead{$m_I$ \tablenotemark{e}} & 
\colhead{$m_H$ \tablenotemark{f}} & 
\colhead{$M$ \tablenotemark{g}} & 
\colhead{Sp. type\tablenotemark{h}} &
\colhead{$V_c$ } &
\colhead{$\mu$\tablenotemark{i}} &
\colhead{$\theta_e$\tablenotemark{j}} & 
}
\startdata
709 &4-402.32&$-0.53$&$2.98$ &$3.03$&0.555&21.68 & 21.34 & $-20.14$ & SB2 & $136 \pm 48$ & 1.106 & 0.30\cr
650 &4-430.0 &$-0.29$&$-4.53$&$4.54$&0.875&23.42 & 23.07 & $-19.71$ & SB2 & $114 \pm 40$ & 1.031 & 0.14\cr 
720 &4-402.31&$0.91$ &$5.29$ &$5.37$ &0.557&21.01 & 20.00 & $-20.75$ & Sbc & $175 \pm 61$ & 1.097 & 0.50\cr
694 &4-493.0 &$6.89$ & $-2.39$&$7.29$&0.849&21.66 & 20.19 & $-21.60$ & El  & $148 \pm 30.$ & 1.035 & 0.25\cr
625 &4-378.0 &$-5.95$& $-5.44$&$8.06$&1.225&24.17 & 23.22 & $-20.18$ & Im  & $138 \pm 48$ & 1.014 & 0.11\cr
653 &4-254.0 &$-14.12$&$1.19$&$14.17$&0.900&22.34 & 21.14 & $-21.19$ & El  & $131 \pm 28$ & 1.013 & 0.18
\enddata

\tablenotetext{a}{ID in the Fern\'andez-Soto, Lanzetta, \& Yahil (1999) catalog.}
\tablenotetext{b}{ID in the Williams et al. (1996) catalog.}
\tablenotetext{c}{Angular distance to the SN 1997ff host in arc seconds.}
\tablenotetext{d}{Spectroscopic redshifts from Cohen et al. (2000)}
\tablenotetext{e}{Apparent AB magnitude in the F814W WFPC2 filter.}
\tablenotetext{f}{Apparent AB magnitude in the F160W NICMOS filter.}
\tablenotetext{g}{Vega-based absolute magnitude $B_J$ for ellipticals, $B$ for
the others.}
\tablenotetext{h}{Spectral type (see text).}
\tablenotetext{i}{Magnification created by this galaxy at the position of the
  host without considering the rest of the lenses in the field.}
\tablenotetext{j}{Einstein radius in arcsec.}

\end{deluxetable}

\begin{figure*}[h]
\epsfig{file=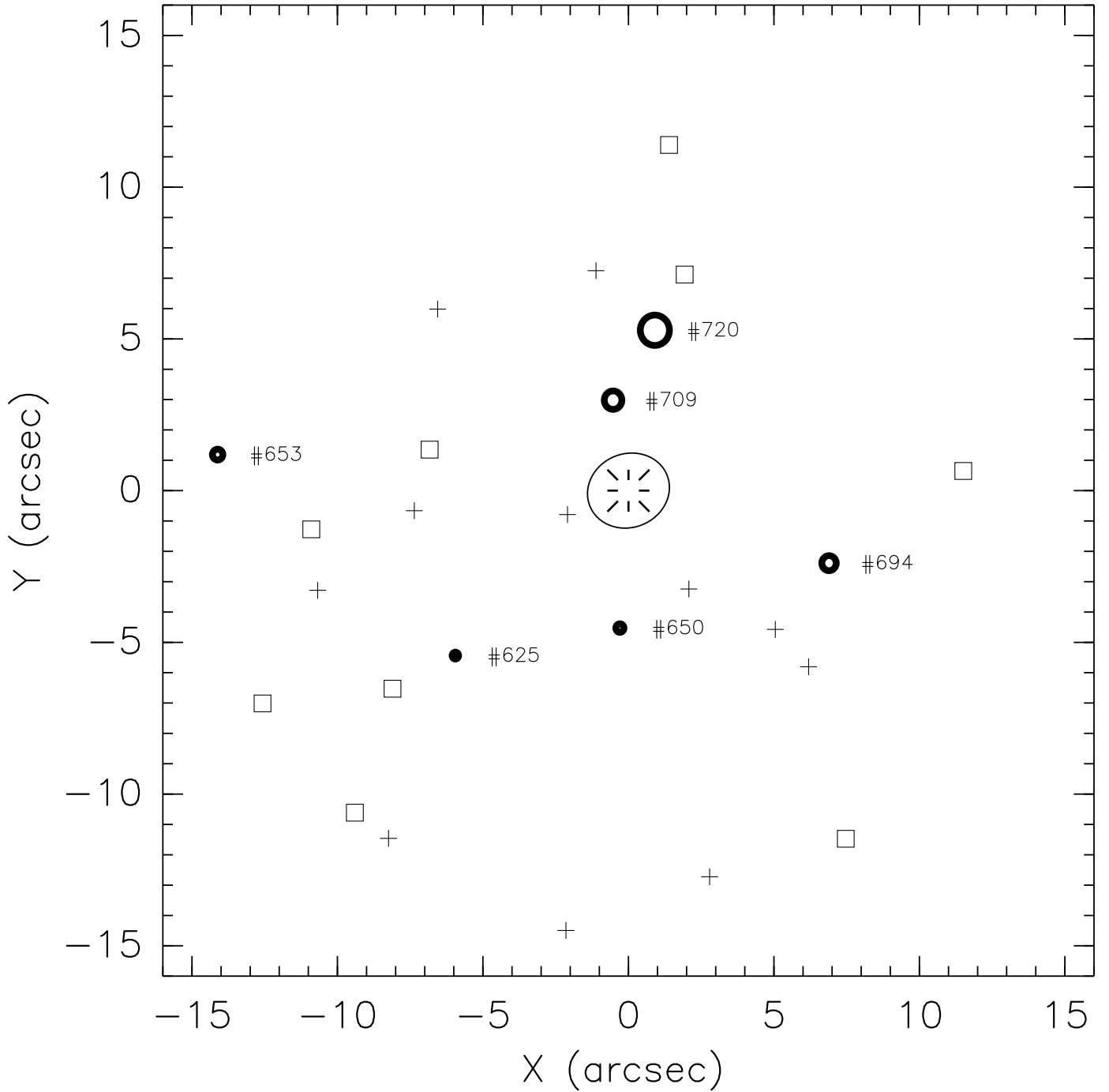}
\caption{The environment of SN 1997ff. The main lenses in Table 1
are represented by circles whose size in proportional to their Einstein radius.
The host galaxy is at the center of the plot, with the shape and orientation
reported in R2001. Foreground galaxies not massive enough to significantly
affect the magnitude of the SN are represented by squares, and crosses
correspond to objects with $z \geq 1.7$. North is up and east is to the left.}
\end{figure*}

\end{document}